\documentclass[12pt]{article}
\usepackage{amsmath,amssymb}
\usepackage[usenames]{color}
\usepackage{pstricks}
\numberwithin{equation}{section}

\newcommand{\nc}{\newcommand}
\def\vvdots{\mathinner{\mkern1mu\raise1pt\vbox{\kern7pt\hbox{.}}\mkern2mu
  \raise4pt\hbox{.}\mkern2mu\raise7pt\hbox{.}\mkern1mu}}
\nc{\fh}{\hat{f}}
\nc{\muh}{\hat{\mu}}
\nc{\nuh}{\hat{\nu}}
\nc{\bib}{\bibitem}
\nc{\al}{\alpha}
\nc{\g}{\gamma}
\nc{\G}{\Gamma}
\nc{\D}{\Delta}
\nc{\eps}{\epsilon}
\nc{\la}{\lambda}
\nc{\La}{\Lambda}
\nc{\var}{\varphi}
\nc{\pa}{\partial}
\nc{\nn}{\nonumber \\ }
\nc{\hf}{\frac{1}{2}}
\nc{\dz}{\frac{dz}{2\pi i}}
\nc{\bin}[2]{\left(\!\!\!\begin{array}{c} {#1}\\ {#2} \end{array}\!\!\!\right)}
\nc{\be}{\begin{equation}}
\nc{\ee}{\end{equation}}
\nc{\bea}{\begin{eqnarray}}
\nc{\eea}{\end{eqnarray}}
\nc{\bra}[1]{\langle {#1}|}
\nc{\ket}[1]{|{#1}\rangle}
\nc{\chit}{\raisebox{0.25ex}{$\chi$}}
\nc{\Db}{\mbox{\boldmath $D$}}
\nc{\Hb}{\mbox{\boldmath $H$}}
\nc{\Hc}{{\cal H}}
\nc{\Rc}{{\cal R}}
\nc{\Lc}{{\cal L}}
\nc{\Vc}{{\cal V}}
\nc{\Ib}{\mbox{\boldmath $I$}}
\nc{\qb}{\bar{q}}
\nc{\ch}{{\rm ch}}
\nc{\R}{{\cal R}}
\nc{\dkk}{\delta_{j,\{k,k'\}}^{(2)}}
\nc{\ddkk}{\delta_{j,\{k,k'\}}^{(4)}}
\nc{\dddkk}{\delta_{j,\{k,k'\}}^{(8)}}
\nc{\dnn}{\delta_{j,\{n,n'\}}^{(2)}}
\nc{\ddnn}{\delta_{j,\{n,n'\}}^{(4)}}
\nc{\dddnn}{\delta_{j,\{n,n'\}}^{(8)}}
\def\vvdots{\mathinner{\mkern1mu\raise1pt\vbox{\kern7pt\hbox{.}}\mkern2mu
  \raise4pt\hbox{.}\mkern2mu\raise7pt\hbox{.}\mkern1mu}}
\nc{\gauss}[2]{\left[\!\!\begin{array}{c} {#1}\\ {#2} \end{array}\!\!\right]}
\nc{\sbin}[2]{\left\{\!\!\!\begin{array}{c} {#1}\\ {#2} 
\end{array}\!\!\!\right\}}
\nc{\sbinlr}[2]{\Big\langle\!\!\begin{array}{c} {#1}\\ {#2} 
\end{array}\!\!\Big\rangle}
\nc{\bino}[2]{\left(\!\!\begin{array}{c} {#1}\\ {#2} \end{array}\!\!\right)}

\definecolor{lightblue}{rgb}{.61,.61,1}
\definecolor{midblue}{rgb}{.7,.7,1}
\definecolor{lightlightblue}{rgb}{.85,.85,1}
\definecolor{lightestblue}{rgb}{.96,.96,1}

\begin{document}

\topmargin -5mm
\oddsidemargin 5mm

\setcounter{page}{1}

\mbox{}\vspace{-16mm}
\thispagestyle{empty}

\begin{center}
{\huge {\bf Polynomial Fusion Rings}}\\[.4cm]
{\huge {\bf of Logarithmic Minimal Models}}

\vspace{7mm}
{\Large J{\o}rgen Rasmussen}\ \ and\ \ {\Large Paul A. Pearce}\\[.3cm]
{\em Department of Mathematics and Statistics, University of Melbourne}\\
{\em Parkville, Victoria 3010, Australia}\\[.4cm]
J.Rasmussen@ms.unimelb.edu.au,\quad P.Pearce@ms.unimelb.edu.au

\end{center}

\vspace{8mm}
\centerline{{\bf{Abstract}}}
\vskip.4cm
\noindent
We identify quotient polynomial rings isomorphic to the recently found fundamental fusion
algebras of logarithmic minimal models. 
\vskip1cm
%
%
\renewcommand{\thefootnote}{\arabic{footnote}}
\setcounter{footnote}{0}

\section{Introduction}
\label{SecIntro}

The fusion algebras of the logarithmic minimal models ${\cal LM}(p,p')$ introduced in \cite{PRZ}
are discussed in \cite{RPperc07,RPfusion07}.
In these works, it is found that closure of the so-called {\em fundamental} fusion algebra 
of ${\cal LM}(p,p')$ requires an infinite set of indecomposable representations of rank 1, 2 or 3.
The former are so-called Kac representations of which some, but in general only some, 
are irreducible (highest-weight) representations. It is recalled that the fundamental fusion
algebra is so named since it is generated from the two fundamental Kac representations
$(2,1)$ and $(1,2)$
\be
 \big\langle (2,1),(1,2)\big\rangle_{p,p'}
\ee
We let $X$ and $Y$ denote the commuting fusion matrices
associated to the representations $(2,1)$ and $(1,2)$, respectively.
Since we consider a countably infinite number 
of representations, $X$ and $Y$ are infinite dimensional.
The main objective of the present work is to establish
the following proposition where $T_n(x)$ and $U_n(x)$ are Chebyshev polynomials
of the first and second kind, respectively, see Appendix \ref{appCheb}.
\\[.2cm]
{\bf Proposition \ref{SecIntro}.1}\ \ \ 
The fundamental fusion algebra of the logarithmic minimal model
${\cal LM}(p,p')$ is isomorphic to the polynomial
ring generated by $X$ and $Y$ modulo
the ideal ${\cal I}_{p,p'}(X,Y)=P_{p,p'}(X,Y)\mathbb{C}[X,Y]$ where
\be
  P_{p,p'}(X,Y)\ =\ \Big(T_p\big(\frac{X}{2}\big)-T_{p'}\big(\frac{Y}{2}\big)\Big)
   U_{p-1}\big(\frac{X}{2}\big)U_{p'-1}\big(\frac{Y}{2}\big)
\label{P}
\ee
that is,
\be
 \big\langle (2,1),(1,2)\big\rangle_{p,p'}\ \simeq\ \mathbb{C}[X,Y]/{\cal I}_{p,p'}(X,Y)
\ee

It is known \cite{Gep91} that a similar isomorphism exists for every {\em rational} conformal
field theory. The proposition above thus extends this to include the {\em irrational}
logarithmic minimal models as well. 
We find, though, that the conjectured existence of an associated
fusion potential in the case of a 
rational conformal field theory \cite{Gep91} does not extend to the irrational ${\cal LM}(p,p')$,
see Appendix \ref{AppFP}.
\subsection*{Notation}
\vskip.1cm
With 
$\mathbb{Z}_{n,m}\ =\ \mathbb{Z}\cap[n,m]$ denoting the set of integers from $n$ to $m$,
both included, we shall be using the following notation:
$a\in\mathbb{Z}_{0,p-1}$; $b\in\mathbb{Z}_{0,p'-1}$; 
$a_0,r_0\in\mathbb{Z}_{1,p-1}$; $b_0,s_0\in\mathbb{Z}_{1,p'-1}$.

\section{Fundamental Fusion Algebra of ${\cal LM}(p,p')$}

A logarithmic minimal model ${\cal LM}(p,p')$ is defined \cite{PRZ} for every coprime pair of
positive integers $p<p'$. The model ${\cal LM}(p,p')$ has central charge
\be
 c\ =\  1-6\frac{(p'-p)^2}{pp'}
\label{c}
\ee
and conformal weights
\be
 \D_{r,s}\ =\ \frac{(rp'-sp)^{2}-(p'-p)^2}{4pp'},\hspace{1.2cm}r,s\in\mathbb{N}
\label{Drs}
\ee

\subsection{Representations}

We recall the set of representations $\{\R_{r,s}^{a,b}\}$ appearing in the 
description of the fundamental fusion algebra of ${\cal LM}(p,p')$ \cite{RPfusion07}.
The representation $\R_{r,s}^{a,b}$ is of rank 1 if $a=b=0$; it is
of rank 2 if $a=0,b\neq0$ or $a\neq0,b=0$; while it is of rank
3 if $a,b\neq0$. The lower indices $r$ and $s$ are positive
integers addressed in the following.

The representations of the form $\R_{r,s}^{0,0}$
are the Kac representations and are also denoted $(r,s)$. 
In connection with the fundamental fusion algebra, there are three
classes of Kac representations: the irreducible, the fully reducible
and the reducible yet indecomposable Kac representations
\be
 \{(r,kp'),(kp,s);\ r\in\mathbb{Z}_{1,p};\ s\in\mathbb{Z}_{1,p'};\ k\in\mathbb{N}\},\ \ \ \ \ 
  \{(kp,k'p');\ k,k'\in\mathbb{N}+1\},\ \ \ \ \ 
  \{(r_0,s_0)\}
\label{Kac}
\ee
here listed in the indicated order. 

The higher-rank representations are classified according to their decomposability.
For $k,k'\in\mathbb{N}$, 
$\R_{pk,s_0}^{a_0,0}$ and $\R_{r_0,p'k'}^{0,b_0}$ are indecomposable representations of rank 2;
$\R_{pk,p'k'}^{a_0,0}$ and $\R_{pk,p'k'}^{0,b_0}$ are indecomposable representations
of rank 2 if $k=1$ or $k'=1$ but decomposable representations of rank 2 if $k,k'>1$;
$\R_{pk,p'k'}^{a_0,b_0}$ is an indecomposable representation of rank 3
if $k=1$ or $k'=1$ but a decomposable representation of rank 3 if $k,k'>1$.

The fully reducibility or decomposability of some of these representations is made
manifest \cite{RPfusion07} by 
\be
 \R_{pk,p'k'}^{a,b}\ =\ \bigoplus_{j=|k-k'|+1,\ \!{\rm by}\ \!2}^{k+k'-1}\R_{pj,p'}^{a,b}
  \ =\ \bigoplus_{j=|k-k'|+1,\ \!{\rm by}\ \!2}^{k+k'-1}\R_{p,p'j}^{a,b},\ \ \ \ \ \ \ k,k'\in\mathbb{N}
\label{kkexp}
\ee
Ensuing identifications are
\be
 \R_{pk,p'k'}^{a,b}\ =\ \R_{pk',p'k}^{a,b}
\label{RequalR}
\ee 
with $(kp,p')=(p,kp')$ corresponding to the identification of a pair of irreducible Kac 
representations of identical conformal weights.
The decompositions (\ref{kkexp}) imply that the fundamental fusion algebra
of ${\cal LM}(p,p')$ can be written in closed form without reference to the fully reducible
Kac representations nor to the decomposable higher-rank representations.
Indeed, according to \cite{RPfusion07},
closure of the fundamental fusion algebra requires the inclusion of the representations 
\be
  \big\langle(2,1), (1,2)\big\rangle_{p,p'}\ =\ \big\langle(r_0,s_0), (pk,s_0), (r_0,p'k), (pk,p'),
   \R_{pk,s_0}^{a_0,0}, \R_{pk,p'}^{a_0,0}, \R_{r_0,p'k}^{0,b_0}, \R_{p,p'k}^{0,b_0}, 
  \R_{pk,p'}^{a_0,b_0}\big\rangle_{p,p'}
\label{A2112}
\ee
where $k\in\mathbb{N}$.

\subsection{Fusion}

The fusion rules of ${\cal LM}(p,p')$ are associative, commutative and separate into
a horizontal and a vertical part \cite{RPfusion07}. 
We indicate this separation with a general but somewhat formal evaluation. Letting 
$A_{r,s}=\bar{a}_{r,1}\otimes\ a_{1,s}$, $B_{r',s'}=\bar{b}_{r',1}\otimes\ b_{1,s'}$,
$\bar{a}_{r,1}\otimes\bar{b}_{r',1}=\bigoplus_{r''}\bar{c}_{r'',1}$ and
$a_{1,s}\otimes b_{1,s'}=\bigoplus_{s''}c_{1,s''}$, our fusion prescription yields
\bea
 A_{r,s}\otimes B_{r',s'}\!\!&=&\!\!\Big(\bar{a}_{r,1}\otimes a_{1,s}\Big)\otimes
  \Big(\bar{b}_{r',1}\otimes b_{1,s'}\Big)
   \ =\ \Big(\bar{a}_{r,1}\otimes\bar{b}_{r',1}\Big)\otimes
  \Big(a_{1,s}\otimes b_{1,s'}\Big)\nn
 \!\!&=&\!\!\Big(\bigoplus_{r''}\bar{c}_{r'',1}\Big)\otimes\Big(\bigoplus_{s''}c_{1,s''}\Big)
  \ =\ \bigoplus_{r'',s''}C_{r'',s''}
\label{rs}
\eea
where $C_{r'',s''}=\bar{c}_{r'',1}\otimes c_{1,s''}$.
As illustration, we have
\be
 \R_{pk,1}^{a,0}\otimes \R_{1,p'k'}^{0,b}\ =\  \R_{pk,p'k'}^{a,b}
\label{kk}
\ee

Since the fundamental fusion algebra is built from repeated fusions of the two
fundamental representations $(2,1)$ and $(1,2)$, we now list all fusions of
one of these fundamental representations with one of the representations 
(all of which are indecomposable) appearing in (\ref{A2112}). In the horizontal
direction, we have 
\bea
 (2,1)\otimes(r_0,s_0) \!\!&=&\!\! (r_0-1,s_0)\oplus (r_0+1,s_0)   \nn
 (2,1)\otimes(pk,s_0) \!\!&=&\!\! \delta_{p,1}\Big((k-1,s_0)\oplus (k+1,s_0)\Big)
    \oplus\big(1-\delta_{p,1}\big)\R_{pk,s_0}^{1,0}\nn
 (2,1)\otimes(r_0,p'k) \!\!&=&\!\! (r_0-1,p'k)\oplus (r_0+1,p'k)   \nn
 (2,1)\otimes(pk,p') \!\!&=&\!\! \delta_{p,1}\Big((k-1,p')\oplus (k+1,p')\Big)
    \oplus\big(1-\delta_{p,1}\big)\R_{pk,p'}^{1,0}
\label{fund21Kac}
\eea
and
\bea
 (2,1)\otimes\R_{pk,s_0}^{a_0,0} \!\!&=&\!\!  
    \delta_{p,2}\Big((2k-2,s_0)\oplus2(2k,s_0)\oplus(2k+2,s_0)\Big)\nn
   &\oplus&\!\! \big(1-\delta_{p,2}\big)\Big(
    \big(1+\delta_{a_0,1}\big)\R_{pk,s_0}^{a_0-1,0}
    \oplus  \big(1-\delta_{a_0,p-1}\big)\R_{pk,s_0}^{a_0+1,0}\nn
   &&\hspace{1.9cm} \oplus\ \delta_{a_0,p-1}\big((pk-p,s_0)\oplus(pk+p,s_0)\big)\Big) \nn
 (2,1)\otimes\R_{pk,p'}^{a_0,0} \!\!&=&\!\!    
    \delta_{p,2}\Big((2k-2,p')\oplus2(2k,p')\oplus(2k+2,p')\Big)\nn
   &\oplus&\!\! \big(1-\delta_{p,2}\big)\Big(
    \big(1+\delta_{a_0,1}\big)\R_{pk,p'}^{a_0-1,0}
    \oplus  \big(1-\delta_{a_0,p-1}\big)\R_{pk,p'}^{a_0+1,0}\nn
   &&\hspace{1.9cm} \oplus\ \delta_{a_0,p-1}\big((pk-p,p')\oplus(pk+p,p')\big)\Big) \nn
 (2,1)\otimes\R_{r_0,p'k}^{0,b_0} \!\!&=&\!\! \R_{r_0-1,p'k}^{0,b_0}\oplus\R_{r_0+1,p'k}^{0,b_0}   \nn
 (2,1)\otimes\R_{p,p'k}^{0,b_0} \!\!&=&\!\!   
   \delta_{p,1}\Big(\R_{1,p'k-p'}^{0,b_0}\oplus\R_{1,p'k+p'}^{0,b_0}\Big)
    \oplus\big(1-\delta_{p,1}\big)\R_{pk,p'}^{1,b_0} \nn
 (2,1)\otimes\R_{pk,p'}^{a_0,b_0} \!\!&=&\!\!    
    \delta_{p,2}\Big(\R_{2k-2,p'}^{0,b_0}\oplus2\R_{2k,p'}^{0,b_0}\oplus\R_{2k+2,p'}^{0,b_0}\Big)\nn
   &\oplus&\!\! \big(1-\delta_{p,2}\big)\Big(
    \big(1+\delta_{a_0,1}\big)\R_{pk,p'}^{a_0-1,b_0}
    \oplus  \big(1-\delta_{a_0,p-1}\big)\R_{pk,p'}^{a_0+1,b_0}\nn
   &&\hspace{1.9cm} 
    \oplus\ \delta_{a_0,p-1}\big(\R_{p,p'k-p'}^{0,b_0}\oplus\R_{p,p'k+p'}^{0,b_0}\big)\Big) 
\label{fund21R}
\eea
while in the vertical direction we have
\bea
 (1,2)\otimes(r_0,s_0) \!\!&=&\!\! (r_0,s_0-1)\oplus (r_0,s_0+1)    \nn
 (1,2)\otimes(pk,s_0) \!\!&=&\!\! (pk,s_0-1)\oplus (pk,s_0+1)         \nn
 (1,2)\otimes(r_0,p'k) \!\!&=&\!\! \R_{r_0,p'k}^{0,1}   \nn
 (1,2)\otimes(pk,p') \!\!&=&\!\! \R_{p,p'k}^{0,1}
\label{fund12Kac}
\eea
and
\bea
 (1,2)\otimes\R_{pk,s_0}^{a_0,0} \!\!&=&\!\!  \R_{pk,s_0-1}^{a_0,0}\oplus \R_{pk,s_0+1}^{a_0,0}  \nn
 (1,2)\otimes\R_{pk,p'}^{a_0,0} \!\!&=&\!\!  \R_{pk,p'}^{a_0,1}  \nn
 (1,2)\otimes\R_{r_0,p'k}^{0,b_0} \!\!&=&\!\!    
    \big(1+\delta_{b_0,1}\big)\R_{r_0,p'k}^{0,b_0-1}
    \oplus  \big(1-\delta_{b_0,p'-1}\big)\R_{r_0,p'k}^{0,b_0+1}\nn
   &\oplus&\!\! \delta_{b_0,p'-1}\big((r_0,p'k-p')\oplus(r_0,p'k+p')\big) \nn
 (1,2)\otimes\R_{p,p'k}^{0,b_0} \!\!&=&\!\!    
    \delta_{p',2}\big((k-1,2)\oplus2(k,2)\oplus(k+1,2)\big)\nn
   &\oplus&\!\! \big(1-\delta_{p',2}\big)\Big(
    \big(1+\delta_{b_0,1}\big)\R_{p,p'k}^{0,b_0-1}
    \oplus  \big(1-\delta_{b_0,p'-1}\big)\R_{p,p'k}^{0,b_0+1}\nn
   &&\hspace{1.9cm} \oplus\ \delta_{b_0,p'-1}\big((pk-p,p')\oplus(pk+p,p')\big)\Big) \nn
 (1,2)\otimes\R_{pk,p'}^{a_0,b_0} \!\!&=&\!\!    
   \big(1+\delta_{b_0,1}\big)\R_{pk,p'}^{a_0,b_0-1}
    \oplus  \big(1-\delta_{b_0,p'-1}\big)\R_{pk,p'}^{a_0,b_0+1}
   \oplus\delta_{b_0,p'-1}\big(\R_{pk-p,p'}^{a_0,0}\oplus\R_{pk+p,p'}^{a_0,0}\big)\nn
\label{fund12R}
\eea
Here we have used that $1\leq p<p'$ and introduced the simplifying notation
\be
 (0,s)\ =\ (r,0)\ =\ \R_{0,s}^{a,b}\ =\ \R_{r,0}^{a,b}\ =\ 0
\ee
Even though we included many details on the fundamental fusion algebras
in \cite{RPperc07,RPfusion07}, the lists (\ref{fund21Kac}) through (\ref{fund12R}) were not 
presented as explicitly as above. 
Finally, it is noted that the Kac representation 
$(1,1)$ is the identity of the fundamental fusion algebra.

\section{Fundamental Fusion Ring of ${\cal LM}(p,p')$}
\label{SecFFR}

\subsection{Fusion Matrices and Fusion Rings}

The fusion algebra, see \cite{DiFMS} for example,
\be
 \phi_i\otimes\phi_j\ =\ \bigoplus_{k\in\mathcal{J}}{{\cal N}_{i,j}}^k\phi_k,\hspace{1cm}i,j\in\mathcal{J}
\ee
of a {\em rational} conformal field theory is finite and
can be represented by a commutative matrix algebra $\big\langle N_i;\ i\in\mathcal{J}\big\rangle$ 
where the entries of the square $|\mathcal{J}|\times|\mathcal{J}|$ matrix $N_i$ are
\be
 {(N_i)_j}^k\ =\ {{\cal N}_{i,j}}^k,\hspace{1cm}i,j,k\in\mathcal{J}
\ee
and where the fusion product $\otimes$ has been replaced by ordinary matrix multiplication.
In \cite{Gep91}, Gepner found that every such algebra is isomorphic to a ring
of polynomials in a finite set of variables modulo an ideal defined as the vanishing conditions
of a finite set of polynomials in these variables. He also conjectured that this ideal of constraints 
corresponds to the local extrema of a potential, see \cite{DiFZ,Aha} for
further elaborations on this conjecture.

Since the fundamental fusion algebra of the logarithmic minimal model
${\cal LM}(p,p')$ contains infinitely many
elements, the associated fusion matrices are infinite-dimensional. The corresponding
conformal field theory is {\em irrational} (in this case {\em logarithmic} \cite{PRZ}) and the results
of Gepner \cite{Gep91} do not necessarily apply. 
We will generally denote these fusion matrices by
$N_{(r,s)}$ or $N_{\R_{r,s}^{a,b}}$, cf. (\ref{A2112}). 
Associativity of the original commutative fusion algebra ensures that these
fusion matrices form a commutative matrix algebra. The fusion matrix associated to the fundamental
representation $(2,1)$ or $(1,2)$ is also denoted $X=N_{(2,1)}$ or $Y=N_{(1,2)}$, respectively.
As we will argue below, every fusion matrix can be written as a polynomial in $X$ and $Y$
and these polynomials are naturally expressed in terms of Chebyshev polynomials, 
see Appendix \ref{appCheb}. With this realization, and in correspondence
with a naive extension of the results by Gepner \cite{Gep91}, we then identify a quotient
polynomial (fusion) ring structure isomorphic to the fundamental fusion algebra of ${\cal LM}(p,p')$.
There does not, on the other hand, appear to be a fusion potential naturally associated to this
fusion ring, see Appendix \ref{AppFP}. 
It is emphasized that this is not in violation of Gepner's results since
our logarithmic minimal model ${\cal LM}(p,p')$ is {\em irrational}.

As preparation for the derivation of the fusion ring, we now turn our attention to 
some relations involving Chebyshev polynomials.

\subsection{Chebyshev Relations}

In the following, we consider two possibly non-invertible and possibly
non-commuting entities $x$ and $y$ and define the polynomial
\be
 M_{p,p'}(x,y)\ =\  U_{2p-1}(x)U_{p'-1}(y)-U_{p-1}(x)U_{2p'-1}(y)
\label{M}
\ee
To ease the notation, 
we will often abbreviate $f(x,y)\equiv g(x,y)$ (mod $M_{p,p'}(x,y)$) simply by $f(x,y)\equiv g(x,y)$.
\\[.2cm] 
{\bf Proposition \ref{SecFFR}.1}\ \ \ For $k\in\mathbb{N}$ and modulo $M_{p,p'}(x,y)$, we have
\be
 U_{pk-1}(x)U_{p'-1}(y)\ \equiv\ U_{p-1}(x)U_{p'k-1}(y)
\label{UUkmodP}
\ee
{\bf Proof}\ \ \ This is trivially true for $k=1,2$ and we use induction in $k$ to
complete the proof. First, though, we prove (\ref{UUkmodP}) for 
$k=3$ in which case
\bea
 U_{3p-1}(x)U_{p'-1}(y)\!\!&=&\!\!\Big(2T_p(x)U_{2p-1}(x)-U_{p-1}(x)\Big)U_{p'-1}(y)\nn
  &\equiv&\!\!2T_p(x)U_{p-1}(x)U_{2p'-1}(y)-U_{p-1}(x)U_{p'-1}(y)\nn
  &=&\!\!U_{2p-1}(x)U_{p'-1}(y)2T_{p'}(y)-U_{p-1}(x)U_{p'-1}(y)\nn
  &\equiv&\!\!U_{p-1}(x)\Big(U_{2p'-1}(y)2T_{p'}(y)-U_{p'-1}(y)\Big)\nn
  &=&\!\!U_{p-1}(x)U_{3p'-1}(y)
\eea
where the three equalities all follow from (\ref{2TU}). The two equivalences are both
immediate consequences of the definition of $M_{p,p'}(x,y)$ in (\ref{M}).
To establish the general induction step for $k\geq3$, we consider 
\bea
 U_{(k+1)p-1}(x)U_{p'-1}(y)\!\!&=&\!\!\Big(2T_p(x)U_{kp-1}(x)-U_{(k-1)p-1}(x)\Big)U_{p'-1}(y)\nn
 &\equiv&\!\!2T_p(x)U_{p-1}(x)U_{kp'-1}(y)-U_{(k-1)p-1}(x)U_{p'-1}(y)\nn
 &=&\!\!2T_p(x)U_{p-1}(x)\Big(2T_{(k-1)p'}(y)U_{p'-1}(y)+U_{(k-2)p'-1}(y)\Big)\nn
  &&-U_{(k-1)p-1}(x)U_{p'-1}(y)\nn
 &\equiv&\!\!U_{p-1}(x)U_{2p'-1}(y)2T_{(k-1)p'}(y)+2T_p(x)U_{(k-2)p-1}(x)U_{p'-1}(y)\nn
 &&-U_{(k-1)p-1}(x)U_{p'-1}(y)\nn
 &=&\!\!U_{p-1}(x)\Big(U_{(k+1)p'-1}(y)-U_{(k-3)p'-1}(y)\Big)\nn
 &&+\Big(U_{(k-1)p-1}(x)+U_{(k-3)p-1}(x)\Big)U_{p'-1}(y)-U_{(k-1)p-1}(x)U_{p'-1}(y)\nn
 &\equiv&\!\!U_{p-1}(x)U_{(k+1)p'-1}(y)
\eea
where, again, all three equalities follow from (\ref{2TU}), 
while the three equivalences follow by induction assumption 
with the second equivalence also relying on (\ref{2TU}).
$\Box$
\\[.2cm]
{\bf Proposition \ref{SecFFR}.2}\ \ \ For $k,k'\in\mathbb{N}$ and modulo $M_{p,p'}(x,y)$, we have
\be
 U_{pk-1}(x)U_{p'k'-1}(y)
  \ \equiv\  \sum_{j=|k-k'|+1,\ \!{\rm by}\ \!2}^{k+k'-1}U_{pj-1}(x)U_{p'-1}(y)
  \ \equiv\ \sum_{j=|k-k'|+1,\ \!{\rm by}\ \!2}^{k+k'-1}U_{p-1}(x)U_{p'j-1}(y)
\label{UUsummodP}
\ee
{\bf Proof}\ \ \ To prove the first equivalence, we initially assume that $k\leq k'$. 
For $k=2n+1$ odd and modulo $M_{p,p'}(x,y)$, we then have
\bea
 U_{p(2n+1)-1}(x)U_{p'k'-1}(y)\!\!&=&\!\!\Big(1+2\sum_{j=1}^nT_{2jp}(x)\Big)U_{p-1}(x)U_{p'k'-1}(y)\nn
 &\equiv&\!\!\Big(1+2\sum_{j=1}^nT_{2jp}(x)\Big)U_{pk'-1}(x)U_{p'-1}(y)\nn
 &=&\!\!\Big(U_{pk'-1}(x)+\sum_{j=1}^n\big(U_{(k'-2j)p-1}(x)+U_{(k'+2j)p-1}(x)\big)\Big)U_{p'-1}(y)
\label{pr32}
\eea
which is readily seen to equal the first sum expression of (\ref{UUsummodP}). 
The first and second equality of (\ref{pr32}) follow from (\ref{UTspec}) and (\ref{2TU}), respectively.
For $k=2n$ even and once again employing (\ref{UTspec}) and (\ref{2TU}), we likewise have
\bea
 U_{2np-1}(x)U_{p'k'-1}(y)\!\!&=&\!\!2\sum_{j=1}^nT_{(2j-1)p}(x)U_{p-1}(x)U_{k'p'-1}(y)\nn
 &\equiv&\!\!2\sum_{j=1}^nT_{(2j-1)p}(x)U_{k'p-1}(x)U_{p'-1}(y)\nn
 &=&\!\!\sum_{j=1}^n\big(U_{(k'+1-2j)p-1}(x)+U_{(k'-1+2j)p-1}(x)\big)U_{p'-1}(y)
\eea
which is also readily seen to equal the first sum expression of (\ref{UUsummodP}).
The first equivalence of (\ref{UUsummodP}) for $k>k'$ follows similarly.
The second equivalence of (\ref{UUsummodP})
is a direct consequence of Proposition \ref{SecFFR}.1.
$\Box$
\\[.2cm]
{\bf Corollary \ref{SecFFR}.3}\ \ \ For $k,k'\in\mathbb{N}$ and modulo $M_{p,p'}(x,y)$, we have
\be
 U_{pk-1}(x)U_{p'k'-1}(y)\ \equiv\ U_{pk'-1}(x)U_{p'k-1}(y)
\label{UUkkmodP}
\ee
{\bf Proof}\ \ \ This follows from Proposition \ref{SecFFR}.2 since the
sum expressions of (\ref{UUsummodP}) are symmetric in $k$ and $k'$.
$\Box$

\subsection{Determination of Fusion Matrices and Ring Structure}

We now show that the generators of the fundamental
fusion algebra (\ref{A2112}) can be expressed as polynomials in the fusion matrices
of the fundamental representations
\be
 X\ =\ N_{(2,1)},\ \ \ \ \ \ \ Y\ =\ N_{(1,2)}
\ee
%
%
{\bf Proposition \ref{SecFFR}.4}\ \ \ Modulo the polynomial $P_{p,p'}(X,Y)$ defined
in (\ref{P}), the matrices
\bea
 N_{(r_0,s_0)}(X,Y)\!\!&=&\!\!U_{r_0-1}\big(\frac{X}{2}\big)U_{s_0-1}\big(\frac{Y}{2}\big)\nn
 N_{(pk,s_0)}(X,Y)\!\!&=&\!\!U_{pk-1}\big(\frac{X}{2}\big)U_{s_0-1}\big(\frac{Y}{2}\big)\nn
 N_{(r_0,p'k)}(X,Y)\!\!&=&\!\!U_{r_0-1}\big(\frac{X}{2}\big)U_{p'k-1}\big(\frac{Y}{2}\big)\nn
 N_{(pk,p')}(X,Y)\!\!&=&\!\!U_{pk-1}\big(\frac{X}{2}\big)U_{p'-1}\big(\frac{Y}{2}\big)
\label{NUKac}
\eea
and
\bea
 N_{{\cal R}_{pk,s_0}^{a_0,0}}(X,Y)\!\!&=&\!\!2T_{a_0}\big(\frac{X}{2}\big)N_{(pk,s_0)}(X,Y)\nn
 N_{{\cal R}_{pk,p'}^{a_0,0}}(X,Y)\!\!&=&\!\!2T_{a_0}\big(\frac{X}{2}\big)N_{(pk,p')}(X,Y)\nn
 N_{{\cal R}_{r_0,p'k}^{0,b_0}}(X,Y)\!\!&=&\!\! 2N_{(r_0,p'k)}(X,Y)T_{b_0}\big(\frac{Y}{2}\big)\nn
 N_{{\cal R}_{p,p'k}^{0,b_0}}(X,Y)\!\!&=&\!\!  2N_{(p,p'k)}(X,Y)T_{b_0}\big(\frac{Y}{2}\big)\nn
 N_{{\cal R}_{pk,p'}^{a_0,b_0}}(X,Y)\!\!&=&\!\! 
   4T_{a_0}\big(\frac{X}{2}\big)N_{(pk,p')}(X,Y)T_{b_0}\big(\frac{Y}{2}\big)
\label{NUR}
\eea
satisfy the fusion rules (\ref{fund21Kac}) through (\ref{fund12R}) with the fusion product 
$\otimes$ and direct summation
$\oplus$ replaced by matrix multiplication and addition, respectively.
Since every participating representation can be written in the form $\R_{r,s}^{a,b}$,
the associated fusion matrix thus reads
\be
 N_{\R_{r,s}^{a,b}}(X,Y)\ =\ \big(2-\delta_{a,0}\big)T_a\big(\frac{X}{2}\big)U_{r-1}\big(\frac{X}{2}\big)
  \big(2-\delta_{b,0}\big)T_b\big(\frac{Y}{2}\big)U_{s-1}\big(\frac{Y}{2}\big)
\label{NR}
\ee
%
%
{\bf Proof}\ \ \ There are 18 fusion rules to establish. The first one appears in (\ref{fund21Kac})
and reads
\bea
 (2,1)\otimes(r_0,s_0)\!\!&\leftrightarrow&\!\! 
  XU_{r_0-1}\big(\frac{X}{2}\big)U_{s_0-1}\big(\frac{Y}{2}\big)
 \ =\ \Big(U_{r_0-2}\big(\frac{X}{2}\big)+U_{r_0}\big(\frac{X}{2}\big)\Big)U_{s_0-1}\big(\frac{Y}{2}\big)\nn
 &\leftrightarrow&\!\! (r_0-1,s_0)\oplus(r_0+1,s_0)
\eea
More generally, the task is to decompose the products
\be
  (2,1)\otimes\R_{r,s}^{a,b}\ \leftrightarrow\  
  X\big(2-\delta_{a,0}\big)T_a\big(\frac{X}{2}\big)U_{r-1}\big(\frac{X}{2}\big)
  \big(2-\delta_{b,0}\big)T_b\big(\frac{Y}{2}\big)U_{s-1}\big(\frac{Y}{2}\big)
\ee
and
\be
  (1,2)\otimes\R_{r,s}^{a,b}\ \leftrightarrow\ 
  \big(2-\delta_{a,0}\big)T_a\big(\frac{X}{2}\big)U_{r-1}\big(\frac{X}{2}\big)
  \big(2-\delta_{b,0}\big)YT_b\big(\frac{Y}{2}\big)U_{s-1}\big(\frac{Y}{2}\big)
\ee
in terms of the polynomials (\ref{NUKac}) and (\ref{NUR}) thereby demonstrating that
the fusion rules (\ref{fund21Kac}) through (\ref{fund12R}) are indeed satisfied. 
To this end, it is noted that
\be
 2P_{p,p'}(X,Y)\ =\ M_{p,p'}\big(\frac{X}{2},\frac{Y}{2}\big)
\label{PM}
\ee
(where $M_{p,p'}(x,y)$ is defined in (\ref{M})) thus permitting us to draw on Proposition \ref{SecFFR}.1 
and Proposition \ref{SecFFR}.2. Establishing the remaining 17 fusion rules 
is now straightforward so we only include one of them as illustration, namely the rule
associated to the fusion product
\be
 (1,2)\otimes\R_{p,p'k}^{0,b_0} \ \leftrightarrow\ U_{p-1}\big(\frac{X}{2}\big)
   2YT_{b_0}\big(\frac{Y}{2}\big)U_{p'k-1}\big(\frac{Y}{2}\big)
\label{12R}
\ee
For $p'=2$, in which case $p=1$ and $b_0=1$, the right side reads
\bea
 \Big(2T_{1}\big(\frac{Y}{2}\big)\Big)^2U_{2k-1}\big(\frac{Y}{2}\big)\!\!&=&\!\!
  U_{2k-3}\big(\frac{Y}{2}\big)+2U_{2k-1}\big(\frac{Y}{2}\big)+U_{2k+1}\big(\frac{Y}{2}\big)\nn
  \!\!&\equiv&\!\!U_{k-2}\big(\frac{X}{2}\big)U_{1}\big(\frac{Y}{2}\big)
    +2U_{k-1}\big(\frac{X}{2}\big)U_{1}\big(\frac{Y}{2}\big)
    +U_{k}\big(\frac{X}{2}\big)U_{1}\big(\frac{Y}{2}\big)\nn
 \!\!&\leftrightarrow&\!\! 
 (k-1,2)\oplus2(k,2)\oplus(k+1,2)
\eea
where the equivalence is modulo $P_{1,2}(X,Y)$. For $p'>2$, the right side of (\ref{12R}) reads
\bea
 U_{p-1}\big(\frac{X}{2}\big)2YT_{b_0}\big(\frac{Y}{2}\big)U_{p'k-1}\big(\frac{Y}{2}\big)
   \!\!&=&\!\!U_{p-1}\big(\frac{X}{2}\big)\Big(
   U_{p'k-b_0-2}\big(\frac{Y}{2}\big)+U_{p'k-b_0}\big(\frac{Y}{2}\big)\nn
  &&\hspace{2cm}
   +\ U_{p'k+b_0-2}\big(\frac{Y}{2}\big)+U_{p'k+b_0}\big(\frac{Y}{2}\big)\Big)
\eea
For $b_0=1$, the right side of this equals
\be
 U_{p-1}\big(\frac{X}{2}\big)\Big(
   U_{p'k-3}\big(\frac{Y}{2}\big)+2U_{p'k-1}\big(\frac{Y}{2}\big)+U_{p'k+1}\big(\frac{Y}{2}\big)\Big)
  \ \leftrightarrow\ 2(p,p'k)\oplus\R_{p,p'k}^{0,2}\ =\ 2\R_{p,p'k}^{0,0}\oplus\R_{p,p'k}^{0,2}
\ee
while for $1<b_0<p'-1$, it equals
\be
 U_{p-1}\big(\frac{X}{2}\big)\Big(
   U_{p'k-b_0-2}\big(\frac{Y}{2}\big)+U_{p'k-b_0}\big(\frac{Y}{2}\big)
   +U_{p'k+b_0-2}\big(\frac{Y}{2}\big)+U_{p'k+b_0}\big(\frac{Y}{2}\big)\Big)
 \ \leftrightarrow\ \R_{p,p'k}^{0,b_0-1}\oplus\R_{p,p'k}^{0,b_0+1}
\ee
whereas it equals
\bea
 && U_{p-1}\big(\frac{X}{2}\big)\Big(
   U_{p'(k-1)-1}\big(\frac{Y}{2}\big)+U_{p'k-(p'-2)-1}\big(\frac{Y}{2}\big)
   +U_{p'k+(p'-2)-1}\big(\frac{Y}{2}\big)+U_{p'(k+1)-1}\big(\frac{Y}{2}\big)\Big)\nn
 &\equiv&\!\! U_{p-1}\big(\frac{X}{2}\big)\Big(
   U_{p'k-(p'-2)-1}\big(\frac{Y}{2}\big)+U_{p'k+(p'-2)-1}\big(\frac{Y}{2}\big)\Big)\nn
  && +\ \Big(U_{p(k-1)-1}\big(\frac{X}{2}\big)+U_{p(k+1)-1}\big(\frac{X}{2}\big)\Big)
     U_{p'-1}\big(\frac{Y}{2}\big)\nn
 &\leftrightarrow&\!\! \R_{p,p'k}^{0,p'-2}\oplus(p(k-1),p')\oplus(p(k+1),p')
\eea
for $b_0=p'-1$ where the equivalence is modulo $P_{1,2}(X,Y)$. This completes the proof
of the fourth fusion rule of (\ref{fund12R}).
$\Box$
\\[.2cm]
{\bf Proposition \ref{SecFFR}.5}\ \ \ The matrices defined by (\ref{NUKac}) and (\ref{NUR}) 
in Proposition \ref{SecFFR}.4 satisfy the fusion
prescription outlined in (\ref{rs}) with the fusion product $\otimes$ and direct summation
$\oplus$ replaced by matrix multiplication and addition, respectively.
\\[.2cm]
{\bf Proof}\ \ \ In analogy with (\ref{rs}) and using (\ref{NR}), we have
\bea
 \R_{r,s}^{a,b}\otimes\R_{r',s'}^{a',b'}\!\!&\leftrightarrow&\!\!
  \left\{\big(2-\delta_{a,0}\big)T_a\big(\frac{X}{2}\big)U_{r-1}\big(\frac{X}{2}\big)
  \big(2-\delta_{b,0}\big)T_b\big(\frac{Y}{2}\big)U_{s-1}\big(\frac{Y}{2}\big)\right\}\nn
 &\times&\!\! \left\{\big(2-\delta_{a',0}\big)T_{a'}\big(\frac{X}{2}\big)U_{r'-1}\big(\frac{X}{2}\big)
  \big(2-\delta_{b',0}\big)T_{b'}\big(\frac{Y}{2}\big)U_{s'-1}\big(\frac{Y}{2}\big)\right\}\nn
 &=&\!\!  \left\{\big(2-\delta_{a,0}\big)T_a\big(\frac{X}{2}\big)U_{r-1}\big(\frac{X}{2}\big)
   \big(2-\delta_{a',0}\big)T_{a'}\big(\frac{X}{2}\big)U_{r'-1}\big(\frac{X}{2}\big)\right\}\nn
 &\times&\!\! \left\{\big(2-\delta_{b,0}\big)T_b\big(\frac{Y}{2}\big)U_{s-1}\big(\frac{Y}{2}\big)
  \big(2-\delta_{b',0}\big)T_{b'}\big(\frac{Y}{2}\big)U_{s'-1}\big(\frac{Y}{2}\big)\right\}\nn
 &=&\!\!
  \left\{\sum_{r'',a''}\big(2-\delta_{a'',0}\big)T_{a''}\big(\frac{X}{2}\big)U_{r''-1}\big(\frac{X}{2}\big)\right\}
  \left\{\sum_{s'',b''}\big(2-\delta_{b'',0}\big)T_{b''}\big(\frac{Y}{2}\big)U_{s''-1}\big(\frac{Y}{2}\big)\right\}\nn
 &\leftrightarrow&\!\!\bigoplus_{r'',s'',a'',b''}\R_{r'',s''}^{a'',b''}
\eea
%
$\Box$\\
In terms of the polynomials (\ref{NUKac}) and (\ref{NUR}) in the commuting variables $X$ and $Y$, 
it is noted that Proposition \ref{SecFFR}.1 corresponds to the identifications
\be
 (kp,p')\ =\ (p,kp')
\ee
of {\em irreducible} Kac representations, while the analogue of the decompositions
(\ref{kkexp}) follow straightforwardly from Proposition \ref{SecFFR}.2 and the product
form of the fusion matrices (\ref{NUR}).
Finally, Corollary \ref{SecFFR}.3 corresponds to the identifications (\ref{RequalR}).
We may thus conclude that, modulo the polynomial $P_{p,p'}(X,Y)$, 
the matrices defined in (\ref{NUKac}) and (\ref{NUR})
provide a fusion-matrix realization of the fundamental fusion algebra of
${\cal LM}(p,p')$.

Our final objective here is to identify the polynomial
ring structure isomorphic to this fusion algebra.
First, we argue that $\mathbb{C}[x,y]$ is equivalent to the span of the combinations
of Chebyshev polynomials (\ref{NUKac}) and (\ref{NUR})
used in the realization of the fundamental fusion algebra.
Since $U_n(z)$ is a polynomial in $z$ of degree $n$, we have
\be
  {\rm span}_{\mathbb{C}}\big\{z^n;\ n\in\mathbb{Z}_{0,N}\big\} \ =\ 
  {\rm span}_{\mathbb{C}}\big\{U_n(z);\ n\in\mathbb{Z}_{0,N}\big\} 
\ee
Furthermore, 
\be
 \big(2-\delta_{a,0}\big)T_a(z)U_{pk-1}(z)\ =\ U_{pk-a-1}(z)+U_{pk+a-1}(z)
\ee
implies that we for $N=\kappa p+\alpha$ where $\kappa\in\mathbb{N}\cup\{0\}$
and $\alpha\in\mathbb{Z}_{0,p-1}$ have
\bea
 {\rm span}_{\mathbb{C}}\big\{z^n;\ n\in\mathbb{Z}_{0,N}\big\}
 \!\!&=&\!\!{\rm span}_{\mathbb{C}}\big\{\big(2-\delta_{a,0}\big)T_a(z)U_{pk-1}(z),\ 
  \big(2-\delta_{a',0}\big)T_{a'}(z)U_{p\kappa-1}(z);\nn
  &&\hspace{4.2cm} k\in\mathbb{Z}_{0,\kappa-1},\ 
  a\in\mathbb{Z}_{0,p-1},\ a'\in\mathbb{Z}_{0,\alpha}\big\}
\eea
For commuting variables $x$ and $y$, we thus have
\bea
 \mathbb{C}[x,y]\!\!&=&\!\! 
   {\rm span}_{\mathbb{C}}\big\{U_n(x)U_{n'}(y);\ n,n'\in\mathbb{N}\cup\{0\}\big\}\nn
 &=&\!\!{\rm span}_{\mathbb{C}}\big\{\big(2-\delta_{a,0}\big)T_a(x)U_{pk-1}(x)
  \big(2-\delta_{b,0}\big)T_{b}(y)U_{p'k'-1}(y);\nn
  &&\hspace{4cm}  k,k'\in\mathbb{N}\cup\{0\},\ 
  a\in\mathbb{Z}_{0,p-1},\ b\in\mathbb{Z}_{0,p'-1}\big\}
\label{Bpp}
\eea

Now, the matrices $X$ and $Y$ satisfy an infinity of conditions, but as demonstrated
above, they are all consequences of a single condition, namely
\be
 P_{p,p'}(X,Y)\ =\ 0
\label{P0}
\ee
We can therefore conclude that the fusion-matrix realization 
of the fundamental fusion algebra of the logarithmic minimal model ${\cal LM}(p,p')$ is isomorphic
to the ring of polynomials in $X$ and $Y$ modulo the ideal defined by (\ref{P0}).
This is the content of our main result, Proposition \ref{SecIntro}.1.

\section{Critical Dense Polymers and Critical Percolation}
\label{SecPP}

When choosing a basis in which to examine the fusion matrices associated to the
fundamental fusion algebra (\ref{A2112}), it is natural to separate the set of generators
into families. First, there is the finite set of reducible
yet indecomposable Kac representations of rank 1. These representations are of the form $(r_0,s_0)$
and there are $(p-1)(p'-1)$ such representations.
The remaining infinitely many representations are of the form $\R_{r,s}^{a,b}$ 
(where $\R_{r,s}^{0,0}=(r,s)$)
and are naturally organized into families labeled by the values of $r,s,a,b$ where $r$ and $s$ are
given modulo $p$ and $p'$, respectively, cf. (\ref{A2112}). An example of such a family is thus 
$\{\R_{pk,s_0}^{p-1,0};\ k\in\mathbb{N}\}$, and every such family is isomorphic to $\mathbb{N}$. 
By simple inspection of (\ref{A2112}), it follows that the number of these infinite-dimensional families is
\be
 N_f\ =\ \frac{(3p-1)(3p'-1)-1}{3}
\ee
This means that the infinite-dimensional fusion matrices are naturally realized as block matrices
where each rectangular block is of dimension $(p-1)(p'-1)\times(p-1)(p'-1)$, 
$(p-1)(p'-1)\times\infty$, $\infty\times(p-1)(p'-1)$
or $\infty\times\infty$, and the total number of blocks is $(N_f+1-\delta_{p,1})^2$.
Multiplication or addition of two matching fusion matrices is performed by first treating the blocks as
entries of $(N_f+1-\delta_{p,1})\times(N_f+1-\delta_{p,1})$-matrices followed by ordinary
multiplication or addition of the matrix blocks as infinite-dimensional matrices.
Once the various blocks have been identified, this arithmetic can of course be carried out
by introducing a common cut-off to the dimensions of the infinite matrix blocks
which is ultimately considered to run off to infinity.

In the two important examples of critical dense polymers ${\cal LM}(1,2)$ 
and critical percolation ${\cal LM}(2,3)$, we will now show that the explicitly constructed
(infinite-dimensional) 
fusion matrices $X$ and $Y$ indeed satisfy the conditions underlying our analysis of the
polynomial fusion ring above, namely $[X,Y]=0$ and $P_{p,p'}(X,Y)=0$.

\subsection{Critical Dense Polymers ${\cal LM}(1,2)$}

In the basis 
\be
 \{(1,1),(2,1),(3,1),\ldots;(1,2),(1,4),(1,6),\ldots;
   {\cal R}_{1,2}^{0,1},{\cal R}_{1,4}^{0,1},{\cal R}_{1,6}^{0,1},\ldots\}
\label{basis12}
\ee
we have
\be
 X  \ =\ \begin{pmatrix} D&0&0 \\ 0&D&0\\  0& 0& D\end{pmatrix},\ \ \ \ \ \ \ 
 Y\ =\ \begin{pmatrix} 0&I&0 \\ 0&0&I\\ 0&D_+& 0\end{pmatrix}
\ee
where each of the 9 entries is an infinite-dimensional square matrix with $D$ and $D_+$ defined
as
\be
 D\ =\ \begin{pmatrix}
0&1&&&\\
1&0&1&&\\
&1&0&1&\\
&&1&0&\\
&&&& \ddots
\end{pmatrix},\hspace{1cm} D_+\ =\ D+2I
\label{D}
\ee
%
%
{\bf Proposition \ref{SecPP}.1}
\be
 [X,Y]\ =\ 0
\label{XY0}
\ee
and
\be
 0\ =\ 2P_{1,2}(X,Y)\ =\ (X-Y^2+2)Y
\label{XY}
\ee
{\bf Proof}\ \ \ With every explicitly written matrix entry being an infinite-dimensional matrix,
the first identity (\ref{XY0}) follows from
\be
 XY\ =\ \begin{pmatrix}
 0&D&0\\ 0&0&D\\ 0&D^2+2D&0\end{pmatrix}\ =\ YX
\ee
while the second identity (\ref{XY}) follows from
\be
 X-Y^2+2\ =\ \begin{pmatrix} D&0&0\\ 0&D&0\\ 0&0&D\end{pmatrix}
  -\begin{pmatrix}0&0&I\\ 0&D_+&0\\ 0&0&D_+\end{pmatrix}
  +\begin{pmatrix}2I&0&0\\ 0&2I&0\\ 0&0&2I\end{pmatrix}
 \ =\ \begin{pmatrix}D_+&0&-I\\ 0&0&0\\ 0&0&0\end{pmatrix}
\label{XY32}
\ee
and hence
\be
 (X-Y^2+2)Y\ =\ \begin{pmatrix}D_+&0&-I\\ 0&0&0\\ 0&0&0\end{pmatrix}
  \begin{pmatrix}0&I&0\\ 0&0&I\\ 0&D_+&0\end{pmatrix}
 \ =\ \begin{pmatrix}0&0&0\\ 0&0&0\\ 0&0&0\end{pmatrix} 
\ee
$\Box$
\\[.2cm]
We emphasize that since $Y$ does not have an inverse, it is obvious that, 
despite the algebraic factorization of $P_{1,2}(X,Y)$, the vanishing
condition (\ref{XY}) for $P_{1,2}(X,Y)$ is inequivalent to the vanishing condition for 
the factor $X-Y^2+2$. This is clear from (\ref{XY32}) as well.

In terms of the fundamental fusion matrices $X$ and $Y$, the fusion matrices 
associated to the representations (\ref{basis12}) read
\be
 (k,1)\ \leftrightarrow\ U_{k-1}\big(\frac{X}{2}\big),\ \ \ \ \ \ \ 
 (1,2k)\ \leftrightarrow\ U_{2k-1}\big(\frac{Y}{2}\big),\ \ \ \ \ \ \ 
  {\cal R}_{1,2k}^{0,1}\ \leftrightarrow\ U_{2k-2}\big(\frac{Y}{2}\big)+U_{2k}\big(\frac{Y}{2}\big)
\ee
where $k\in\mathbb{N}$.
According to Proposition \ref{SecIntro}.1, the
fundamental fusion algebra of critical dense polymers
${\cal LM}(1,2)$ is isomorphic to the quotient polynomial
ring structure generated by $X$ and $Y$ modulo
the ideal defined by (\ref{XY}). Abbreviating the ideal by its defining polynomial, we thus have 
\be
 \big\langle (1,2),(2,1)\big\rangle_{1,2}\ \simeq\ \mathbb{C}[X,Y]/\big(XY-Y^3+2Y\big)
\ee

\subsection{Critical Percolation ${\cal LM}(2,3)$}

In the basis 
\bea
 &&\{(1,1),(1,2);
  (2,1),(4,1),(6,1),\ldots; 
  (2,2),(4,2),(6,2)\ldots;
  (1,3),(1,6),(1,9),\ldots;\nn
 &&\hspace{0.5cm}
  (2,3),(4,3),(6,3),\ldots;
  {\cal R}_{2,1}^{1,0},{\cal R}_{4,1}^{1,0},{\cal R}_{6,1}^{1,0},\ldots;
  {\cal R}_{2,2}^{1,0},{\cal R}_{4,2}^{1,0},{\cal R}_{6,2}^{1,0},\ldots;
  {\cal R}_{2,3}^{1,0},{\cal R}_{4,3}^{1,0},{\cal R}_{6,3}^{1,0},\ldots;\nn
 &&\hspace{0.5cm}
  {\cal R}_{1,3}^{0,1},{\cal R}_{1,6}^{0,1},{\cal R}_{1,9}^{0,1},\ldots;
  {\cal R}_{2,3}^{0,1},{\cal R}_{2,6}^{0,1},{\cal R}_{2,9}^{0,1},\ldots;
  {\cal R}_{1,3}^{0,2},{\cal R}_{1,6}^{0,2},{\cal R}_{1,9}^{0,2},\ldots;
  {\cal R}_{2,3}^{0,2},{\cal R}_{2,6}^{0,2},{\cal R}_{2,9}^{0,2},\ldots;\nn
 &&\hspace{0.5cm}
  {\cal R}_{2,3}^{1,1},{\cal R}_{4,3}^{1,1},{\cal R}_{6,3}^{1,1},\ldots;
  {\cal R}_{2,3}^{1,2},{\cal R}_{4,3}^{1,2},{\cal R}_{6,3}^{1,2},\ldots \}
\label{basis23}
\eea
we have
\be
 X\ =\ \left(\begin{array}{ccccccccccccccc}
  0&E_u&E_d&0&0&0&0&0&0&0&0&0&0&0\\
 0&0&0&0&0&I&0&0&0&0&0&0&0&0\\
 0&0&0&0&0&0&I&0&0&0&0&0&0&0\\
 0&0&0&0&I&0&0&0&0&0&0&0&0&0\\
 0&0&0&0&0&0&0&I&0&0&0&0&0&0\\
 0&D_+&0&0&0&0&0&0&0&0&0&0&0&0\\
 0&0&D_+&0&0&0&0&0&0&0&0&0&0&0\\
 0&0&0&0&D_+&0&0&0&0&0&0&0&0&0\\
 0&0&0&0&0&0&0&0&0&I&0&0&0&0\\
 0&0&0&0&0&0&0&0&0&0&0&0&I&0\\
 0&0&0&0&0&0&0&0&0&0&0&I&0&0\\
 0&0&0&0&0&0&0&0&0&0&0&0&0&I\\
 0&0&0&0&0&0&0&0&0&D_+&0&0&0&0\\
 0&0&0&0&0&0&0&0&0&0&0&D_+&0&0      \end{array}\right)
\label{X23}
\ee
and
\be
 Y\ =\ \left(\begin{array}{ccccccccccccccc}
  \Gamma&0&0&E_d&0&0&0&0&0&0&0&0&0&0\\
 0&0&I&0&0&0&0&0&0&0&0&0&0&0\\
 0&I&0&0&I&0&0&0&0&0&0&0&0&0\\
 0&0&0&0&0&0&0&0&I&0&0&0&0&0\\
 0&0&0&0&0&0&0&0&0&I&0&0&0&0\\
 0&0&0&0&0&0&I&0&0&0&0&0&0&0\\
 0&0&0&0&0&I&0&I&0&0&0&0&0&0\\
 0&0&0&0&0&0&0&0&0&0&0&0&I&0\\
 0&0&0&2I&0&0&0&0&0&0&I&0&0&0\\
 0&0&0&0&2I&0&0&0&0&0&0&I&0&0\\
 0&0&0&D&0&0&0&0&I&0&0&0&0&0\\
 0&0&0&0&D&0&0&0&0&I&0&0&0&0\\
 0&0&0&0&0&0&0&2I&0&0&0&0&0&I\\
 0&0&0&0&0&0&0&D&0&0&0&0&I&0      \end{array}\right)
\label{Y23}
\ee
where
\be
 \Gamma\ =\ \begin{pmatrix}0&1\\ 1&0\end{pmatrix},\hspace{1cm}
  E_u\ =\ \begin{pmatrix} 1&0&0&\ldots\\ 0&0&0&\ldots \end{pmatrix},
  \hspace{1cm} E_d\ =\ \begin{pmatrix} 0&0&0&\ldots\\ 1&0&0&\ldots \end{pmatrix}
\ee
Denoting the explicitly written entries of $X$ (and similarly of $Y$) by $X_{i,j}$
where $i,j\in\mathbb{Z}_{1,14}$, the entry $X_{1,1}$ is a $2\times2$-matrix; the entry
$X_{1,j}$ for $j\in\mathbb{Z}_{2,14}$ consists of 2 infinite rows; the entry
$X_{i,1}$ for $i\in\mathbb{Z}_{2,14}$ consists of 2 infinite columns; whereas
the entry $X_{i,j}$ for $i,j\in\mathbb{Z}_{2,14}$ is an
infinite-dimensional matrix like (\ref{D}).
\\[.2cm]
{\bf Proposition \ref{SecPP}.2}
\be
 [X,Y]\ =\ 0
\ee
and
\be
 0\ =\ 2P_{2,3}(X,Y)\ =\ X\big(X^2-Y^3+3Y-2\big)\big(Y^2-1\big)
\label{XY23}
\ee
{\bf Proof}\ \ \ As in the case of Proposition \ref{SecPP}.1, this proposition follows by direct inspection.
$\Box$
\\[.2cm]
In terms of the fundamental fusion matrices $X$ and $Y$, the fusion matrices 
associated to the representations (\ref{basis23}) read
\bea
& (1,1)\ \leftrightarrow\ 1,\hspace{.5cm}
 (1,2)\ \leftrightarrow\ Y,\hspace{.5cm} 
 (2k,1)\ \leftrightarrow\ U_{2k-1}\big(\frac{X}{2}\big),\hspace{.5cm} 
 (2k,2)\ \leftrightarrow\ U_{2k-1}\big(\frac{X}{2}\big)Y   &\nn
& (1,3k)\ \leftrightarrow\ U_{3k-1}\big(\frac{Y}{2}\big),\hspace{.5cm}  
 (2k,3)\ \leftrightarrow\ U_{2k-1}\big(\frac{X}{2}\big)\big(Y^2-1\big) &\nn
&  \R_{2k,1}^{1,0}\ \leftrightarrow\ XU_{2k-1}\big(\frac{X}{2}\big),    \hspace{.5cm}
 \R_{2k,2}^{1,0}\ \leftrightarrow\ XU_{2k-1}\big(\frac{X}{2}\big)Y, \hspace{.5cm}
  \R_{2k,3}^{1,0}\ \leftrightarrow\ XU_{2k-1}\big(\frac{X}{2}\big)\big(Y^2-1\big)  &\nn  
&  \R_{1,3k}^{0,1}\ \leftrightarrow\ YU_{3k-1}\big(\frac{Y}{2}\big),   \hspace{.5cm}
   \R_{1,3k}^{0,2}\ \leftrightarrow\ \big(Y^2-2\big)U_{3k-1}\big(\frac{Y}{2}\big) &\nn
& \R_{2,3k}^{0,1}\ \leftrightarrow\  XYU_{3k-1}\big(\frac{Y}{2}\big),   \hspace{.5cm}
  \R_{2,3k}^{0,2}\ \leftrightarrow\  X\big(Y^2-2\big)U_{3k-1}\big(\frac{Y}{2}\big)  &\nn
& \R_{2k,3}^{1,1}\ \leftrightarrow\ XU_{2k-1}\big(\frac{X}{2}\big)\big(Y^3-Y\big)  ,\hspace{.5cm}
   \R_{2k,3}^{1,2}\ \leftrightarrow\  XU_{2k-1}\big(\frac{X}{2}\big)\big(Y^4-3Y^2+2\big)    &
\eea
where $k\in\mathbb{N}$.
According to Proposition \ref{SecIntro}.1, the
fundamental fusion algebra of critical percolation
${\cal LM}(2,3)$ is isomorphic to the quotient polynomial
ring structure generated by $X$ and $Y$ modulo
the ideal defined by (\ref{XY23}). Abbreviating the ideal by its defining polynomial, we thus have 
\be
 \big\langle (1,2),(2,1)\big\rangle_{2,3}\ \simeq\ \mathbb{C}[X,Y]/
  \big(X^3Y^2-X^3-XY^5+4XY^3-2XY^2-3XY+2X\big)
\ee

\section{Conclusion}

We have derived a fusion-matrix realization of the fundamental fusion algebra 
\cite{RPperc07,RPfusion07} of every logarithmic minimal model ${\cal LM}(p,p')$ \cite{PRZ}. 
The various fusion matrices are all expressed in terms of
Chebyshev polynomials in the two infinite-dimensional fundamental fusion matrices $X$ and $Y$
corresponding to the fundamental representations $(2,1)$ and $(1,2)$, respectively.
In terms of this realization, we have identified the quotient polynomial
ring structure isomorphic to
the fundamental fusion algebra itself. This extends the regime of validity of Gepner's result \cite{Gep91}
on the existence of such a quotient polynomial
ring isomorphic to a rational conformal field theory to the irrational
logarithmic minimal models. We have found, though, that the conjectured existence of an 
associated polynomial fusion potential \cite{Gep91} does not extend to the logarithmic minimal models.
We have worked out explicit realizations of the fundamental fusion matrices in the cases
of critical dense polymers ${\cal LM}(1,2)$ and critical percolation ${\cal LM}(2,3)$, and hence
of the full fusion-matrix realizations of the associated fundamental fusion algebras.
We have verified that these explicit matrices satisfy the basic constraints underlying our
construction of the fusion rings.

The fundamental fusion algebras presented in \cite{RPperc07,RPfusion07} are supported,
within a lattice formulation, by extensive numerical studies of associated integrable
lattice models. Despite the vastness of this numerical data set, the fusion rules can only
be considered conjectural. It is therefore very reassuring that the fusion algebra is
isomorphic to a polynomial
fusion ring whose ideal is defined by a single vanishing condition 
which, in turn, corresponds to the natural identification of the
two irreducible highest-weight representations $(2p,p')$ and $(p,2p')$ of identical conformal weights.
\vskip.5cm
\subsection*{Acknowledgments}
\vskip.1cm
\noindent
This work is supported by the Australian Research Council. PAP thanks Giuseppe Mussardo 
and Aldo Delfino for hospitality at the Statistical Physics Section 
of SISSA and Luc Frappat and Eric Ragoucy for hospitality at LAPP.

\appendix

\section{Chebyshev Polynomials}
\label{appCheb}

\subsection{Chebyshev Polynomials of the First Kind}

Recursion relation:
\be
 T_{n}(x)\ =\ 2xT_{n-1}(x)-T_{n-2}(x),\ \ \ \ \ \ \ n=2,3,\ldots
\label{Trec}
\ee
Initial conditions:
\be
 T_0(x)\ =\ 1,\ \ \ \ \ \ \ T_1(x)\ =\ x
\ee
Examples:
\bea
 T_2(x)\!\!&=&\!\! 2x^2-1\nn
 T_3(x)\!\!&=&\!\! 4x^3-3x\nn
 T_4(x)\!\!&=&\!\! 8x^4-8x^2+1\nn
 T_5(x)\!\!&=&\!\! 16x^5-20x^3+5x
\label{Tex}
\eea

\subsection{Chebyshev Polynomials of the Second Kind}

Recursion relation:
\be
 U_{n}(x)\ =\ 2xU_{n-1}(x)-U_{n-2}(x),\ \ \ \ \ \ \ n=2,3,\ldots
\label{Urec}
\ee
Initial conditions:
\be
 U_0(x)\ =\ 1,\ \ \ \ \ \ \ U_1(x)\ =\ 2x
\ee
Examples:
\bea
 U_2(x)\!\!&=&\!\!4x^2-1\nn
 U_3(x)\!\!&=&\!\!8x^3-4x\nn
 U_4(x)\!\!&=&\!\!16x^4-12x^2+1\nn
 U_5(x)\!\!&=&\!\!32x^5-32x^3+6x
\label{Uex}
\eea
Extension:
\be
 U_{-1}(x)\ =\ 0
\ee
Decomposition of product:
\be
 U_m(x)U_n(x)\ =\ \sum_{j=|m-n|,\ \!\!{\rm by}\ \!2}^{m+n}U_j(x)
\label{UUU}
\ee

\subsection{Relating Chebyshev Polynomials of the First and Second Kind}

Basic relation:
\be
 2T_n(x)\ =\ U_n(x)-U_{n-2}(x),\ \ \ \ \ \ \ n\in\mathbb{N}
\label{TU}
\ee
Generalization of the basic relation (\ref{TU}):
\be
 2T_{n}(x)U_{m-1}(x)\ =\ \begin{cases}
   U_{n+m-1}(x)-U_{|n-m|-1}(x),\hspace{1cm}&n>m\\[.2cm]
   U_{n+m-1}(x),  &n=m\\[.2cm]
   U_{n+m-1}(x)+U_{|n-m|-1}(x),  &n<m
 \end{cases}
\label{2TU}
\ee
Applying (\ref{TU}) and (\ref{UUU}) in the given order to the left side of (\ref{2TU}) yields a difference of
two sums which simplifies to the right side of (\ref{2TU}).
\\[.2cm]
Special expansions, with $p\in\mathbb{N}$:
\bea
 U_{(2n+1)p-1}(x)\!\!&=&\!\!\Big(1+2\sum_{j=1}^nT_{2jp}(x)\Big)U_{p-1}(x)\nn
 U_{2np-1}(x)\!\!&=&\!\!2\sum_{j=1}^nT_{(2j-1)p}(x)U_{p-1}(x)
\label{UTspec}
\eea
These relations follow by induction in $n$. In particular, the induction step used in establishing the first
relation reads
\bea
 U_{(2n+1)p-1}(x)\!\!&=&\!\!2T_{2np}(x)U_{p-1}(x)+U_{(2n-1)p-1}(x)\nn
 &=&\!\!2T_{2np}(x)U_{p-1}(x)+\Big(1+2\sum_{j=1}^{n-1}T_{2jp}(x)\Big)U_{p-1}(x)\nn
  &=&\!\!\Big(1+2\sum_{j=1}^nT_{2jp}(x)\Big)U_{p-1}(x)
\eea
where the first equality is a consequence of (\ref{2TU}). The second relation in (\ref{UTspec})
follows similarly.
\\[.2cm]
Derivative:
\be
 \partial_x T_n(x)\ =\ nU_{n-1}(x),\ \ \ \ \ \ \ n\in\mathbb{N}\cup\{0\}
\ee

\section{Fusion Potential}
\label{AppFP}

For $1\leq p<p'$, we now show that the single constraint $M(x,y)=0$, where $M(x,y)=M_{p,p'}(x,y)$
is defined in (\ref{M}) with $[x,y]=0$, cannot be derived from a polynomial potential $V(x,y)$ as the
condition defining the local extrema of $V(x,y)$.
The conditions for local extrema imply that the partial derivatives of $V(x,y)$ must 
vanish modulo $M(x,y)$. Also, since
$M(x,y)$ must be generated from $V(x,y)$, the former must be expressible as
a linear combination of the partial derivatives of the latter. 
We can thus characterize the polynomial potential $V(x,y)$ by the conditions
\bea
 &\partial_x V(x,y)\ =\ f(x,y)M(x,y),\hspace{1cm} \partial_y V(x,y)\ =\ g(x,y)M(x,y)&\nn
 &\alpha\partial_x V(x,y)+\beta\partial_y V(x,y)\ =\ M(x,y)&
\eea
for some $\alpha,\beta\in\mathbb{C}$ and polynomials $f(x,y)$ and $g(x,y)$.
We have four situations, depending on $\alpha$ and $\beta$ being 0 or not,
all of which we now discard one by one.
Assuming $\alpha=\beta=0$, we are immediately faced with the contradiction $M(x,y)=0$.
It is noted that since $M(x,y)$ is asymmetric in its dependence on $x$ and $y$ due to 
the inequality $p<p'$, the two cases $\alpha=0,\beta\neq0$ and $\alpha\neq0,\beta=0$
should be treated separately. 

Assuming $\alpha\neq0, \beta=0$, we integrate $\alpha\partial_x V(x,y)=M(x,y)$ to obtain
\be
 V(x,y)\ =\ \frac{1}{\alpha}\Big(\frac{1}{2p}T_{2p}(x)U_{p'-1}(y)-\frac{1}{p}T_p(x)U_{2p'-1}(y)\Big)
  +\bar{V}(y)
\ee
for some polynomial $\bar{V}(y)$, thus implying
\be
 \frac{1}{\alpha}\Big(\frac{1}{2p}T_{2p}(x)U'_{p'-1}(y)-\frac{1}{p}T_p(x)U'_{2p'-1}(y)\Big)
  +\bar{V}'(y)\ =\ g(x,y)\big(U_{2p-1}(x)U_{p'-1}(y)-U_{p-1}(x)U_{2p'-1}(y)\big)
\ee
Considering this as an identification of polynomials in $x$ with focus on the leading terms, we find that
\be
 \frac{1}{2p\alpha}\big(2^{2p-1}x^{2p}\big)\big(2^{p'-1}(p'-1)y^{p'-2}\big)+\ldots\ =\ 
 g(x,y)\big(2^{2p-1}x^{2p-1}\big)\big(2^{p'-1}y^{p'-1}\big)+\ldots
\ee
Matching these for $g(x,y)$ polynomial (in $y$, in particular)
requires $p'=1$ and $g(x,y)=0$, but $1\leq p<p'$.

Assuming $\alpha=0, \beta\neq0$, we likewise obtain the requirement $p=1$ and $f(x,y)=0$.
This implies $\partial_x V(x,y)=0$ and $\partial_y V(x,y)=xU_{p'-1}(y)-U_{2p'-1}(y)$.
Integrating the latter with respect to $y$ yields a potential $V(x,y)$ with non-trivial dependence
on $x$ in contradiction with $\partial_x V(x,y)=0$.

Assuming $\alpha\neq0, \beta\neq0$, polynomial identification yields $\alpha f(x,y)+\beta g(x,y)=1$
and we are left with the two conditions
\be
 \partial_x V(x,y)\ =\ f(x,y)M(x,y),\hspace{1cm}
  \partial_y V(x,y)\ =\ \frac{1}{\beta}\big(1-\alpha f(x,y)\big)M(x,y)
\ee
We compute the double derivatives
\bea
 \partial_y\partial_x V(x,y)\!\!&=&\!\! \partial_y f(x,y)M(x,y)+f(x,y)\partial_y M(x,y)\nn
 \partial_x\partial_y V(x,y)\!\!&=&\!\! -\frac{\alpha}{\beta}\partial_x f(x,y)M(x,y)
   +\frac{1}{\beta}\big(1-\alpha f(x,y)\big)\partial_x M(x,y)
\label{dd}
\eea
If $f(x,y)=0$, we have $\partial_y\partial_x V(x,y)=0$ and
$\partial_x\partial_y V(x,y)=(1/\beta)\partial_x M(x,y)\neq0$ so $f(x,y)\neq0$. From (\ref{dd}), 
we read off the bounds 
\bea
 {\rm deg}_x\big[\partial_y\partial_x V(x,y)\big]\!\!&\leq&\!\!  
   {\rm deg}_x f(x,y)+{\rm deg}_x M(x,y)\nn
 {\rm deg}_x\big[\partial_x\partial_y V(x,y)\big]\!\!&\leq&\!\!
   {\rm deg}_x f(x,y)+{\rm deg}_x M(x,y)-1
\label{ineq}
\eea
where ${\rm deg}_x h(x,y)$ denotes the degree of $h(x,y)$ as a polynomial in $x$.
An inconsistency is thus reached if the first bound is saturated. From
\be
 \partial_y\partial_x V(x,y)\ =\ U_{2p-1}(x)\partial_y\big[f(x,y)U_{p'-1}(y)\big]
   -U_{p-1}(x)\partial_y\big[f(x,y)U_{2p'-1}(y)\big]
\ee
and the expansion $f(x,y)=x^{d_f}f_0(y)+\ldots$ where $d_f={\rm deg}_x f(x,y)$ (such that
$f_0(y)\neq0$ since $f(x,y)\neq0$), we conclude that saturation of the first inequality
(\ref{ineq}) is prevented if and only if $\partial_y\big[f_0(y)U_{p'-1}(y)\big]=0$.
Since $p'>1$, the polynomial $U_{p'-1}(y)$ is non-constant implying the sought contradiction
$f_0(y)=0$.

Considering (\ref{PM}), 
we thus conclude that the conjectured existence of a polynomial fusion potential in the case of a 
{\em rational} conformal field theory \cite{Gep91} does {\em not} 
extend to the {\em irrational} ${\cal LM}(p,p')$.

It is noted that having fewer polynomial
conditions (here only $M(x,y)=0$) than variables (here $x$ and $y$) is not enough
to prevent a polynomial potential from existing.  
A single polynomial condition given by a function of
$\alpha x+\beta y$ only, for example, can be easily integrated to yield the desired potential.
It was the particular `semi-factorized' form of the single condition 
$M(x,y)=0$ above which allowed us to exclude the possibility of a polynomial potential.


\end{document}